# Conductive Paintable 2D Layered $MoS_2$ Inks

Elaine Carroll[1], Darragh Buckley[1], David McNulty[1], and Colm O'Dwyer[1,2,z]

[1]School of Chemistry, University College Cork, Tyndall National Institute, and Environmental Research Institute, Cork, T12 YN60, Ireland

[2]AMBER@CRANN, Trinity College Dublin, Dublin 2, Ireland

Corresponding Author: [z E-mail: c.odwyer@ucc.ie]

Conductive and paintable inks of 2D layered $MoS_2$ with aspect ratio-dependent conductivity are demonstrated. Using ultrasonically assisted solvent-exfoliation of $MoS_2$, high concentration 2D and few-layer suspensions become inks that provide coherent films when painted. Conductivity of paintable 2D $MoS_2$ inks can be modulated by length and width, where the conductivity is linked to the painting direction. Reducing the painted film width, increases conductivity for similar length, and the films conductivity is aspect ratio-dependent. Inks of solvent-exfoliated 2D $MoS_2$ can be painted without polymeric additives.

Research on two-dimensional (2D) layered materials such as graphene(1), transition metal dichalcogenides (TMDs), carbide and carbonitride relatives called MXenes,(2) has progressed beyond discovery to myriad applications.(3-5) In all applications where electronic transport is important, 2D materials(6), their assemblies, deposition, homo- and hetero-interfaces(7, 8), layered networks, and composite with polymers(9), strongly influence the nature of conductivity and consequently, the type of electrode or device(10). Photonics and optoelectronic investigations(11) are also sensitive to engineering of 2D materials and their composites.(12)

Of the TMD class of 2D materials, $MoS_2$ remains popular, relatively easy to prepare, and chemically stable. It is a classic layered material(13) and solid-state lubricant in its bulk form. It has been examined extensively in 2D electronic and optoelectronics(14-17) due to the relative ease in many exfoliation methods. It can also be grown using physical and vapor deposition as single molecular layers on various substrates, and more recently by electrochemical methods.(18)

Exfoliation of parent bulk crystals to give 2D and few-layer $MoS_2$ dispersions in solvents may also be promising functional inks of various types. Whether thixotropic or dilatant, inks from particle dispersions are the basis of many printing and coating technologies(19, 20). Incorporating new functionality using 2D materials with standard processing has potential for interesting applications. (21)

Inspired by $MoS_2$ used in semi-fluid thixotropic lubricants, we developed a 2D material paint without using polymer additives from high concentration suspensions. Using ultrasonically assisted solvent exfoliation, useful for high-through and high quality dispersions of graphene and a range of TMDs.(4, 5, 22, 23), conductive $MoS_2$ paints are reported here.

Modulating the electrical properties of $MoS_2$ and many assemblies or networks of deposited TMD flakes is sensitive to the method of deposition and growth method of the 2D materials. Conductive polymer additives can hide underlying properties by adding sheet-to-sheet grain boundary and contact resistance, and this is important for paintable inks of 2D materials that may display conductivity that is dependent on a coating method for example(24).

Here, we developed inks using high concentrations of exfoliated $MoS_2$ 2D layer suspensions that can be painted as uniform films on substrates. Conductivity is controlled at material-level by high quality exfoliated 2D sheets. Painted $MoS_2$ in-plane electronic conductivity is altered by the painting direction and aspect ratio of coated films. This 2D material paint and insights into the correlation between paint application, aspect ratio and electrical conductivity has scope for diverse application requiring TMD and layers of 2D materials as a cohesive and conductive thin films that can be painted into uniform layers or patterns.

**Experimental**

$MoS_2$ powder with as-received average particle size < 2μm (Aldrich) was used as received. $MoS_2$ was subsequently exfoliated in solvent using ultrasonication into 2D and few-layer MoS$_2$. To create inks, we followed a procedure like our work on 2D Bi$_2$Te$_3$.(24) A mass of 0.5 g of MoS$_2$ was sonicated in 1 mL of 1-cyclohexenyl pyrrolidine (CHP) solvent for 800 mins. The 2D MoS$_2$ suspension was then centrifuged at 4500 rpm for 45 min. For each dispersion, the supernatant was removed and the remaining mixture (ink) was then painted onto substrates (glass, Si and SiO$_2$) using an artist brush. The material was distributed evenly across the glass slides in several aspect ratios (width, length) to constant thickness of ~200 μm. The substrates were then annealed at 100 °C for 3 h. Physical characterization was carried out using scanning electron microscopy (SEM) using a Hitachi S4800, transmission electron microscopy (TEM) at 200 kV using a JEOL 2100F, atomic force microscopy (AFM) using a Park XE-100 in non-contact mode, Raman scattering spectroscopy, and electrical transport measurements in 2-probe geometry using a Keithley 2612B SourceMeter and a probe station.

**Results and Discussion**

Painted, conductive 2D layered MoS$_2$ films with different aspect ratio ($W \times L$) are shown in Fig. 1(a), shown schematically in Fig. 1(b). Using SEM, the morphology of the paint is shown in Fig. 1(c) and we note a relatively smooth, coherent deposition that does not contain voids or holes during painting or after drying. These films contain only exfoliated MoS$_2$, no polymer or conductive additives to modify viscosity or conductivity. AFM measurements of average thickness and roughness for a typical high aspect ratio painted film. Statistical analysis indicated an average consistent thickness of ~200 μm with roughness varying from 400-650 nm (rms) for a film painted using an artist brush. In Fig. 2, we show single 2D MoS$_2$ (a,b) and few-layered MoS$_2$ (c,d) morphology after solvent exfoliation. Widening of the van der Waals spacing from solvent intercalation and buckling are consistently observed for few-layer MoS$_2$. Raman scattering spectra confirm high quality MoS$_2$ post exfoliation and (Fig. 2(e)), and MoS$_2$ 2D sheets exhibit an expected shift in optical phonon modes, and the higher frequency $E^2_{1u}$ phonon compared to bulk MoS$_2$. Normally, Coulombic interactions between adjacent layers cause Davydov splitting that defines the frequency difference between the ($E^2_{1u}$, $E^1_{2g}$) conjugate pair. The $E^2_{1u}$ mode represents sulfur atoms on adjacent layers moving in phase and exfoliation appears to modify terminating sulfur bonds that alter this mode. The ratio of $E^1_{2g}$ to $A_{1g}$ remains relatively unchanged, in agreement with the layered structure we see by TEM.

We investigated the painting of the MoS$_2$ inks onto glass substrates that were cleaned by UV-ozone treatment. Two-probe transport using flat-bottom spherical In-Ga eutectic ohmic contacts across several painted strips from the MoS$_2$ ink with various aspect ratio are show in in Figure 3. Ohmic response is found in all cases, and roughly doubles in conductivity as the length is halved (Figure 4(a-c)). The seemingly simplistic ohmic behavior appears to follow a length-dependent resistivity interpretation. We posit that tortuosity and complex minimum conductance paths are possibly formed when the width (perpendicular to the 2D material painted direction) is increased. The average in-plane conductivity is ~1.2 – 1.6×10$^{-5}$ S cm$^{-1}$ for these 2D MoS$_2$ painted films, analyzed along the direction of painting. As the aspect ratio is reduced to a square ($L = W$), the lateral 2D and few-layer sheet-to-sheet contacts provide alternative conduction paths between MoS$_2$ or likely around scattering centers. The absolute conductance of wider films is evidently lower than for narrower ones (Fig. 4(a)), the conductivity slightly reduces to ~1.5×10$^{-5}$ S cm$^{-1}$. A similar relative trend holds as a function of overall painted film area (Fig. 4(b)). When we analyze this using a linear trend, the conductivity varies as 2.8×10$^{-5}$ S cm$^{-1}$·cm$^{-1}$ ($W \times L$). Thus, widening the 2D MoS$_2$ paint provides conductance paths that improve overall conductivity measured along the length ($L$). Using the basic definition of resistance

$$R = \rho \frac{L}{Wt} \quad \text{(Eq. 1)}$$

where $\rho$ is resistivity and $t$ is the film thickness, and can be written in terms of conductance, $S$ as

$$S = \frac{1}{R_s}\left(\frac{W \times L}{L^2}\right) \quad \text{(Eq. 2)}$$

where $W \times L$ is the area and $\mathbf{R_s} = \frac{\rho}{t}$ is the film sheet resistance. When $L = W$, the sheet resistance and measured resistance are formally equivalent. For a $L = W = 1.5$ cm, the conductance is $2.32 \times 10^{-5}$ S, giving a sheet resistance of 43.1 kΩ and a corresponding conductivity of $1.55 \times 10^{-5}$ S cm$^{-1}$, therefore predicting a thickness $t \sim 150$ μm. For standard length-dependent resistance, changes to film conductance be modulating the width is accommodated within the definition in Eq. 2 such that scaling according to $\frac{S}{W} = \frac{1}{R_s}\left(\frac{1}{L}\right)$ or $\boldsymbol{S \times L^2} = \frac{1}{R_s}(\boldsymbol{W \times L})$ should yield consistent values (from the slope) for sheet resistance $R_s$. These plots are shown in Figs 4(c) and 4(d), and in both cases we note difference in the values of $R_s$ due to changes in width, and thus aspect ratio. Sheet resistance reduces from 26.3 kΩ to 7.7 kΩ when the width is reduced vs total painted length (Fig. 4(c)) and similarly, we note a reduction from 21.5 kΩ to 3.04 kΩ for narrower 2D MoS$_2$ painted films vs total painted area.

This is confirmed in Fig. 4(e), where we find that lower aspect ratios (*W/L*) provide more conductive films using 2D MoS$_2$ analyzed along the painted direction. As aspect ratio is dimensionless, the conductivity is higher for any aspect ratio when the width (parallel to source-drain) is narrower. This behavior may be particular to layered materials where effects such as shear-thinning or texturization from 2D sheet unidirectional orientation can occur during painting to influence in-plane electronic conductivity. A common figure or merit for conductive ink deposits that assesses electronic networking that influence conductivity is the conductivity-concentration product FOM = $\sigma C$. In this case, with a 500 mg mL$^{-1}$ 2D MoS$_2$ nominal concentration, the FOM is 0.0075 S cm$^{-1}$ · mg mL$^{-1}$ as a lower bound. The state of the art for graphene is ~6000, and so modification of resistivity from underlying texture or grain-to-grain contacts are more readily accessible in the less conductive, uncompressed painted MoS$_2$ inks in this case. Other deposition methods might also impart a directional conductivity dependence(25). Wider films introduce tortuosity to the grain boundary resistivity for given thickness, and the nature of the grain-boundary contacts influence conductivity.

Fundamental transport models for materials with narrow bandgaps may explain conductivity with 3D vs 2D reduced dimensionality, and future studies underway will address some general queries in painted films or assemblies with random sheet-to-sheet contacts: does the density of contacts between 2D sheets (for a given thickness) control overall conductivity? Is the lowest resistance path a function of preferred directionality of 2D sheets, and how does compression alter porosity and interconnection to affect conductivity (26)? Conductivity measured along the width (*W*) is similarly modified by variation of the length

(*L*), and so can anisotropy in 2D material deposition, variation in 2D sheet anisotropy or thickness(27), affect the global anisotropy in conductivity(28) for different painted films of 2D materials we show here? For any functional, complex shaped films, electrodes or related where the 2D material is printed, painted, sprayed or deposited in a manner that adds 'texture' to the material's nanostructure, or preferential orientation of 2D and few-layer sheets, the dependence of conductivity on film aspect ratio is intrinsically important.

**Conclusions**

Smooth, paintable films from 2D $MoS_2$ ink are possible using solvent exfoliation without polymer additives. Electrical conductivity is enhanced by depositing narrower films at most aspect ratios, or shorter films at fixed width. Future work is necessary to understand the fundamental basis for reduced conductivity in wider painted films of equal length, and the prospects of paintable 2D materials in applications where electronic conductivity modification is important. The ability to paint arbitrary coatings provides a method for printable electronics, complex patterned catalytic, electrochemical or other coatings/electrodes on a wide range of substrates.

**Acknowledgments**

We acknowledge the support of the Irish Research Council (IRC) with an Enterprise Partnership Scheme with Analog Devices B.V. under award EPSPG/2011/160 and from a Government of Ireland Postgraduate Award under contract GOIPG/2014/206. Support from Science Foundation Ireland under grant no 14/IA/2581 is also acknowledged.

**Figure 1**

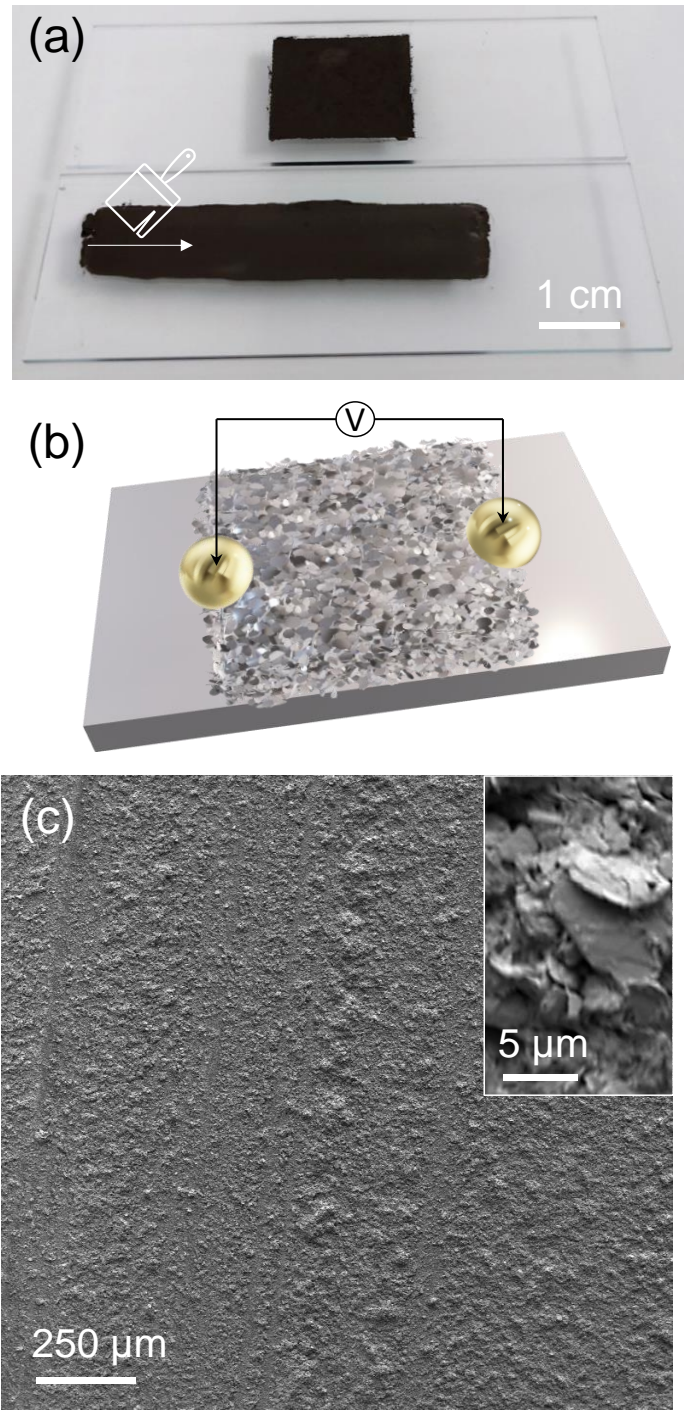

**Figure 1.** (a) Image of a typical painted film from 2D $MoS_2$ ink on a standard 75 mm × 25 mm glass substrate. (b) Schematic representation of 2-probe measurement of painted 2D $MoS_2$ film using flat-bottom In-Ga eutectic ohmic contact droplets of ~4 mm² contacted area wetted to the $MoS_2$ surface. (c) Plan view SEM image of the top surface of the 2D $MoS_2$ ink as a painted film.

**Figure 2**

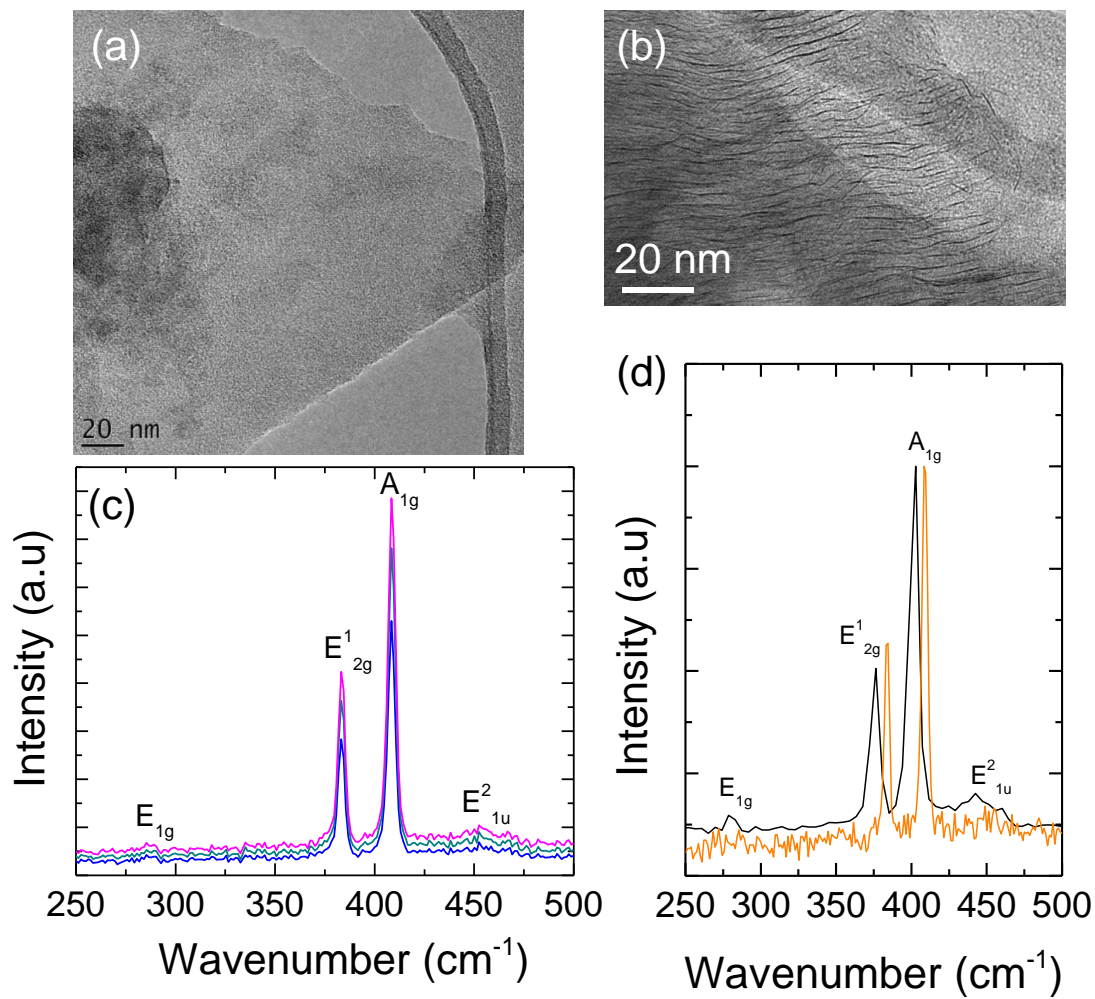

**Figure 2.** (a) Bright field plan-view TEM image of a single 2D $MoS_2$ flake and (b) cross-sectional TEM image of exfoliated few-layer $MoS_2$. (c) Raman scattering spectra of exfoliated 2D $MoS_2$ ink suspensions. (d) Raman scattering spectra for bulk $MoS_2$ powder and exfoliated $MoS_2$.

**Figure 3**

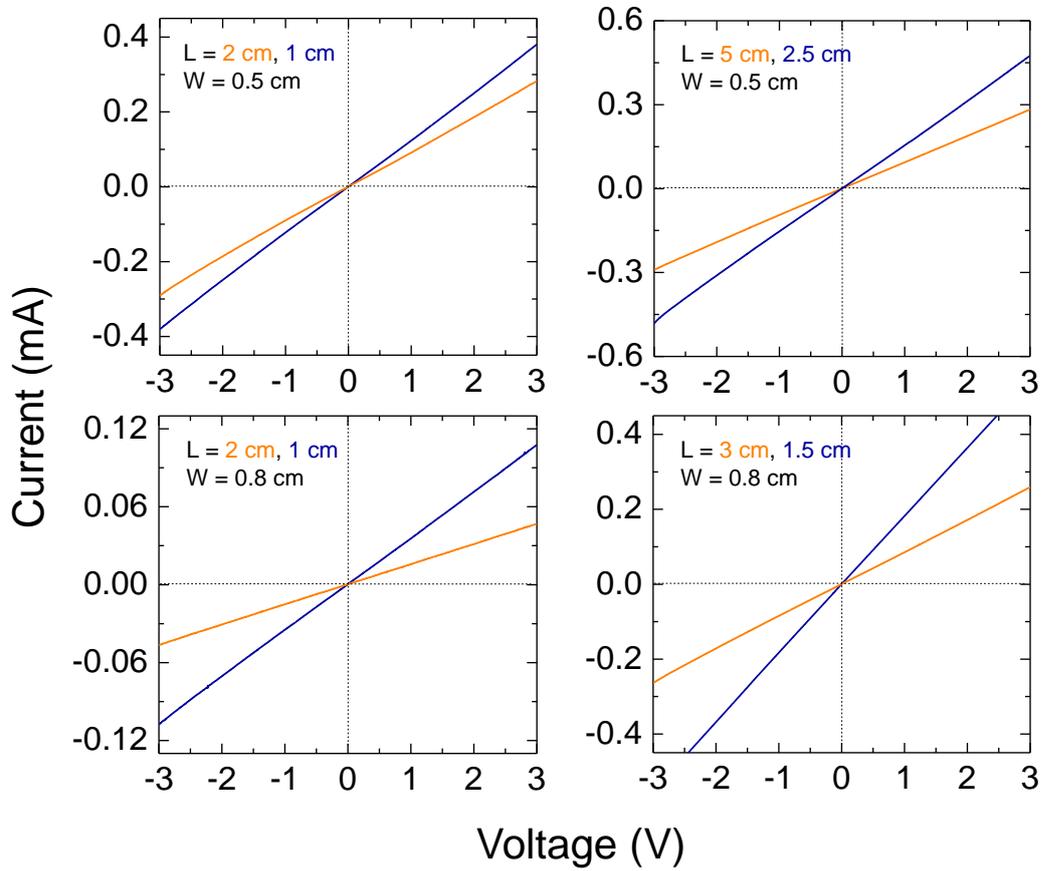

**Figure 3.** I-V response of a set of painted films of exfoliated MoS$_2$ ink on glass, with different length and width (see insets).

**Figure 4**

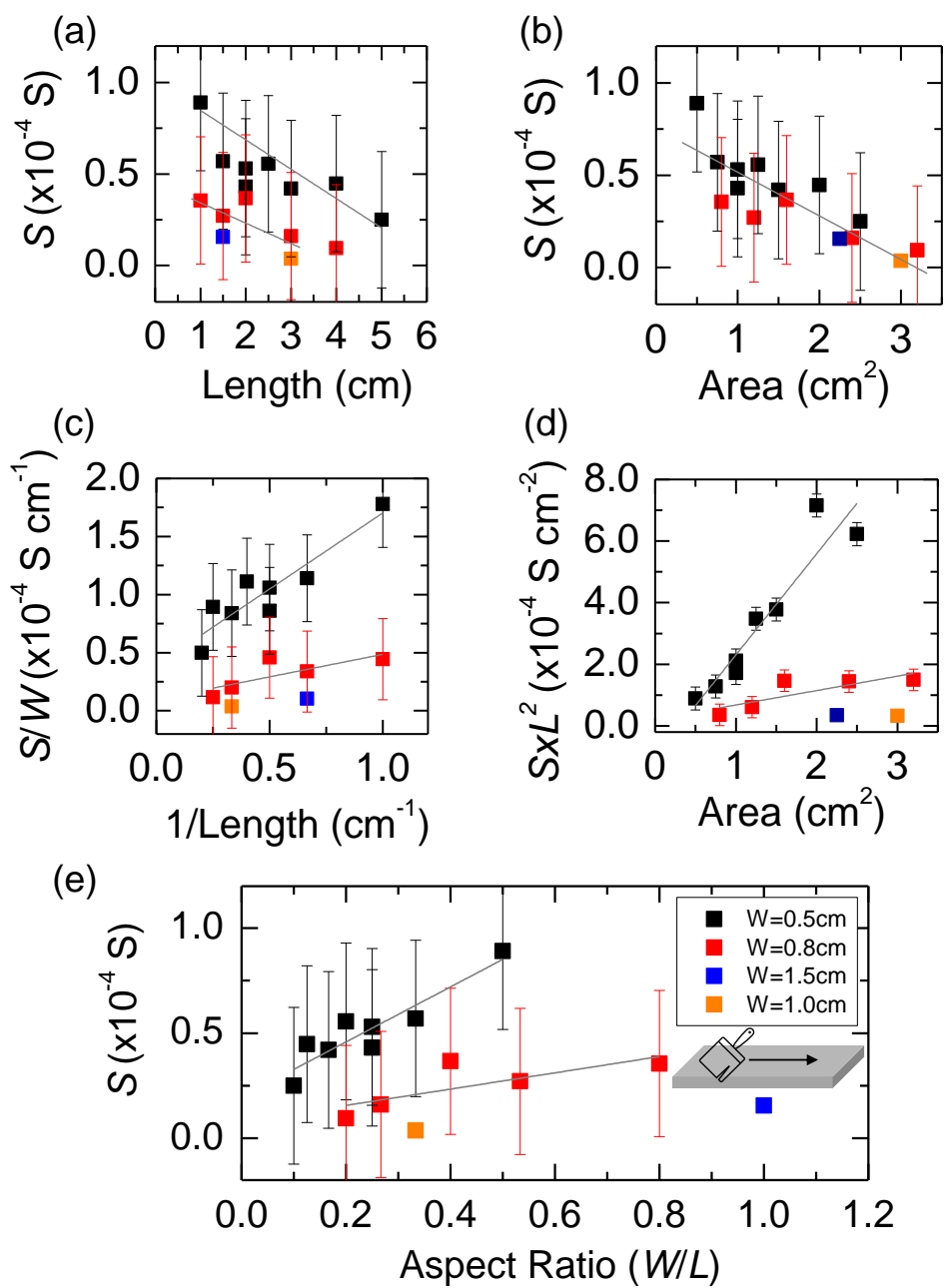

**Figure 4.** Conductance of several painted MoS$_2$ films of various length ($L$) and width ($W$) as a function of (a) painted film length and (b) total film area ($W \times L$). In (c), the conductance data from (a) is scaled by film width according to $\frac{S}{W} = \frac{1}{R_s}\left(\frac{1}{L}\right)$ and in (d) the areal conductance is scaled as $S \times L^2 = \frac{1}{R_s}(W \times L)$ to also give the slope as $R_s$. (e) Conductance as a function of aspect ratio ($W/L$), which directly yields a slope of $R_s$ following Eq, 2. The conductance was measured along the direction of the painting direction, i.e. $L$.